\documentclass[conference]{IEEEtran}
\usepackage{romannum}
\usepackage{amsmath}
\usepackage{graphicx}
\usepackage{multirow}
\usepackage{amssymb,bm,bbm}
\usepackage{algpseudocode}
\usepackage{algorithm}
\usepackage{array}
\usepackage{subfigure}
\usepackage{enumerate}

\usepackage{ifthen}
\usepackage{cite}
\usepackage{mathtools}
\usepackage{multicol}
\ifCLASSINFOpdf
\else

\fi

\begin{document}

\title{Deep Learning Assisted User Identification in Massive Machine-Type Communications}

\author{\IEEEauthorblockN{Bryan Liu\IEEEauthorrefmark{1}, Zhiqiang Wei\IEEEauthorrefmark{1}, Jinhong Yuan\IEEEauthorrefmark{1}, and Milutin Pajovic\IEEEauthorrefmark{2}\\
\IEEEauthorrefmark{1}School of Electrical Engineering and Telecommunications, the University of New South Wales\\
\IEEEauthorrefmark{2}Mitsubishi Electric Research Laboratories, Cambridge, USA\\
Email: bryan.liu@unsw.edu.au}}

\maketitle

\begin{abstract}
In this paper, we propose a deep learning aided list approximate message passing (AMP) algorithm to further improve the user identification performance in massive machine type communications.
A neural network is employed to identify a \emph{suspicious device} which is most likely to be falsely alarmed during the first round of the AMP algorithm.
The neural network returns the false alarm likelihood and it is expected to learn the unknown features of the false alarm event and the implicit correlation structure in the quantized pilot matrix.
Then, via employing the idea of list decoding in the field of error control coding, we propose to enforce the suspicious device to be inactive in every iteration of the AMP algorithm in the second round.
The proposed scheme can effectively combat the interference caused by the suspicious device and thus improve the user identification performance.
Simulations demonstrate that the proposed algorithm improves the mean squared error performance of recovering the sparse unknown signals in comparison to the conventional AMP algorithm with the minimum mean squared error denoiser.
\end{abstract}
\IEEEpeerreviewmaketitle

\section{Introduction}
Triggered by explosive applications of Internet-of-Things (IoT), massive machine-type communications (mMTC), where a large number of sensors are envisioned to transmit short messages sporadically \cite{3gpp,Zhiqiang,WeiJSTSP2019}, have become one of the dominant communication paradigms for future wireless networks.
In order to accommodate such a massive connectivity, grant-free random access schemes were proposed and have got industry and academia consensus on its applicability for mMTC\cite{SunPNC,Liu18ma,RAbbas18}.
By contrast to grant-based schemes, in grant-free access schemes, each device directly transmits its pilot and payload data in one shot, once it has a transmission demand \cite{Liu18ma}.
In addition, limited by the channel coherence time, it is impossible to allocate orthogonal pilot sequences to massive IoT devices \cite{Liu18ma}.
The employment of non-orthogonal pilot sequences among devices yields the device activity detection and channel estimation as a critical but challenging task\cite{SunJUICE,SunJUICE2019}.
Fortunately, the sparsity in device activity enables the possibility to tackle the aforementioned issues via employing compressed sensing (CS) techniques \cite{Liu18ma}.

Most recently, one of CS techniques, approximate message passing (AMP) with minimum mean squared error (MMSE) denoiser has been employed to identify the device activity \cite{WeiYu}, via recasting it as a sparse linear inverse problem and exploiting the statistics of the wireless channels.
The authors \cite{Liu18} proved the remarkeable performance gain in the user identification via employing the AMP algorithm combined with massive multiple-input multiple output (MIMO) technique.
However, the AMP-based user identification highly relies on the assumption of independent and identically distributed (i.i.d.) Gaussian distributed pilot sequences of each device.
In practice, the IoT devices usually equip a low-resolution digital-to-analog converter (DAC) to save its cost and power consumption \cite{lowADC}, which makes the i.i.d. Gaussian distributed pilot sequence idealistic or unapproachable.
In addition, this assumption implies that the correlation structure among pilot sequences has not been exploited by AMP, which means there is potential to further improve the user identification performance.

Deep learning has emerged as a powerful tool to further augment the existing wireless communication technologies \cite{Milutin2019,DLAMP_Philip2016, DLISTA, DLFISTA}.
Furthermore, deep learning has been applied for conventional CS algorithms \cite{DLAMP_Philip2016, DLISTA, DLFISTA}, which resulted in significantly improved accuracy and reduced computational complexity.
Particularly, in \cite{DLAMP_Philip2016}, it was shown that the iterations of the AMP algorithm can be unfolded into several layers of neural network.
After training the constructed neural network, the ``learned AMP'' provides increased accuracy.
Moreover, in \cite{DLAMP}, a learned denoising-based AMP (LDAMP) network was proposed.
By replacing the denoiser in the AMP algorithm with a neural network, the LDAMP has shown an enhanced performance compared to the conventional denoising-based AMP (D-AMP) \cite{DBAMP}.
In summary, the aforementioned algorithms utilize deep learning as a tool to optimize the system performance based on a certain training metric, which is commonly the normalized mean squared error (NMSE).
Therefore, deep learning has a potential to improve the user identification performance via learning the underlying correlation structure among quantized pilot sequences in mMTC.

In this paper, we propose a deep learning based false alarm likelihood (FAL) estimator to assist conventional AMP algorithm for user identification and channel estimation.
In particular, the neural network is designed and trained to estimate the FAL of each device, based on the observations at the receiver and the estimates attained by an AMP algorithm.
Then, comparing all the obtained FALs, we can find the \emph{suspicious device} which is most likely to be falsely alarmed.
Via employing the idea of list decoding from the field of error control coding \cite{OSD}, we propose to restart the AMP algorithm with enforcing the suspicious device as inactive.
Therefore, we name the proposed scheme as deep learning assisted list AMP (DL-LAMP) algorithm.
It is worth noting that compared with finding the most likely miss-detected device and forcing it to be active, finding the suspicious device and enforcing it to be inactive is a simpler approach that does not require any channel state information.
Simulation results show that the proposed DL-LAMP algorithm provides up to 0.8 dB performance gain of NMSE at a signal-to-noise ratio (SNR) of 40 dB, compared to the conventional AMP-MMSE algorithm.

\section{Preliminaries}
\subsection{System Model}
Assume $N$ potential devices transmit packets to a common receiver through a multiple access channel.
Each device has a transmission probability Pr$(a_n=1)=\rho$, $0 \leq \rho \leq 1$.
Here, $a_n$ denotes the activity of device $n$, where $a_n \in \{0,1\}$.
Specifically, $a_n=0$ indicates that device $n$ is inactive and the device is active if $a_n=1$.
We use $K$ to denote the number of active devices.
Any active device will transmit a packet which contains a complex $M$-length pilot sequence $\mathbf{u}_n \in \mathbb{C}^{M \times 1}$.
Each element $u_{m,n}$ in the pilot matrix $\mathbf{U}$ is chosen by $u_{m,n} \sim \mathcal{CN}(0,\frac{1}{M})$, $m \in \{1,2,...,M\}$ and $n \in \{1,2,...,N\}$, where $\mathcal{CN}(0,\frac{1}{M})$ denotes the circularly symmetric complex Gaussian distribution with zero mean and variance $\frac{1}{M}$.
Consider the complex channel fading coefficient $h_n \in \mathbb{C}$ from device $n$ to the receiver at a time slot as $h_n \sim \mathcal{CN}(0,1)$. We have $x_n=a_nh_n$ which captures the joint effect of device activity and channel fading.

The sensors in mMTC are assumed to be low-cost, battery-limited, and thus can only equip a low-resolution DAC.
Hence, a quantized pilot matrix is further considered.
Define an element-wise quantization function $\mathcal{Q}(\cdot, b)$ with reference to the first argument, where $b$ indicates the number of quantization bits.
We consider a uniform quantization function $\mathcal{Q}(\cdot, b)$ which has $2^b$ discrete output levels with equal step between adjacent output levels.
Since each entry in the pilot matrix obeys a complex Gaussian distribution with a variance of $\frac{1}{M}$, the quantization outputs are assumed in the range $[-\frac{3}{\sqrt{2M}},+\frac{3}{\sqrt{2M}}]$ for the real and imaginary components, respectively, which covers approximately $99.7\%$ of the pilot realization \cite{threesigma}.
The pilot realization out of this range would be saturated to the upper bound and lower bound correspondingly.
The quantization function follows the mid-riser rule \cite{Quantization}.
Given a complex value $\kappa=\kappa_r+(\kappa_\jmath)\jmath$, $\mathcal{Q}(\cdot, b)$ returns a quantized complex value by
\begin{equation}\label{Quantization}
  \mathcal{Q}({\kappa}, b) = \Delta \cdot\bigg{(}\lfloor\frac{\kappa_r}{\Delta}+\frac{1}{2}\rfloor\bigg{)} + \Delta \cdot\bigg{(}\lfloor\frac{\kappa_\jmath}{\Delta}+\frac{1}{2}\rfloor\bigg{)}\jmath,
\end{equation}
where $\Delta=\frac{6}{\sqrt{2M}} \times \frac{1}{2^b-1}$.
As a result, the quantized pilot sequence of device $n$ is defined as $\mathbf{p}_n=\mathcal{Q}(\mathbf{u}_n,b)$ and the received signal during pilot transmission is obtained as:
\begin{gather}\label{Ax}
\mathbf{y}=\sum_{n=1}^{N}\mathbf{p}_na_nh_n+\mathbf{z} = \sum_{n=1}^{N}\mathbf{p}_nx_n+\mathbf{z} = \mathbf{Px}+\mathbf{z},
\vspace{-5mm}
\end{gather}
where $\mathbf{z}\in \mathbb{C}^{M \times 1}$ refers to the additive white Gaussian noise (AWGN) with each entry $z_m \sim \mathcal{CN}(0,\sigma_z^2)$.
The pilot matrix $\mathbf{P}=[\mathbf{p}_1,\mathbf{p}_2,...,\mathbf{p}_N] \in \mathbb{C}^{M \times N}$ gathers the pilot sequences of the devices and the unknown vector $\mathbf{x}=[x_1,x_2,...,x_N]^T \in \mathbb{C}^{N \times 1}$ collects the variables $x_n$ to be recovered.
The receiver identifies the users' activity and estimates their channels, i.e., estimates the unknown vector $\mathbf{x}\in \mathbb{C}^{N \times 1}$, based on the observation $\mathbf{y} \in \mathbb{C}^{M \times 1}$ and the pilot matrix $\mathbf{P}$.
This leads to a under-determined problem due to the fact that $M \ll N$.
In addition, since the number of active devices is much less than that of potential devices, i.e., $K \ll N$, CS techniques, such as AMP, can be employed to solve this problem.

\subsection{Overview of AMP Algorithm}
First proposed in \cite{AMP}, AMP has been broadly investigated for solving the sparse linear inverse problems.
%
%
As a low computational complexity algorithm, the AMP algorithm performs iterative updates to recover the sparse unknown signals $\mathbf{x}$.
Define $\mathbf{v}^t \in \mathbb{C}^{M \times 1}$ as the residual errors between the observations $\mathbf{y}$ and the corresponding signals of the estimates $\mathbf{\hat{x}}^t = [\hat{x}_1^t, \hat{x}_2^t,..., \hat{x}_N^t]^{T}$ in the $t$-th iteration.
Then, by initializing $\mathbf{\hat{x}}^0 = \mathbf{0}$ and $\mathbf{v}^0 = \mathbf{y}$, the AMP algorithm mainly comprises of two steps of updates:
\begin{align}
\mathbf{\hat{x}}^{t+1} &= \eta(\mathbf{P}^*\mathbf{v}^t+\mathbf{\hat{x}}^{t}, \sigma_h, t), \label{x_hat}\\
\mathbf{v}^{t+1} &= \mathbf{y} - \mathbf{P}\mathbf{\hat{x}}^{t+1} + \frac{N}{M}\mathbf{v}^t\langle \eta'(\mathbf{P}^*\mathbf{v}^t+\mathbf{\hat{x}}^{t}, \sigma_h, t)\rangle\label{v}
\end{align}
where $\langle \cdot \rangle$ denotes the empirical averaging operation, $\eta(\cdot)$ refers to the denoiser function, $\eta'(\cdot)$ expresses the first-order derivative of $\eta(\cdot)$ with respect to the first argument, $\mathbf{P}^*$ indicates the Hermitian transpose of matrix $\mathbf{P}$, and $\sigma_h$ denotes the standard deviation of the channel fading coefficient.

The denoiser in the AMP algorithm plays an important role for reducing the estimation error and maintaining the sparsity of the unknown vector $\mathbf{x}$.
The MMSE denoiser exploits the prior distribution of the unknown vector $\mathbf{x}$ and thus outperforms the well-known soft thresholding denoiser \cite{STLASSO}. In this paper, we consider the AMP algorithm with the MMSE denoiser.
In particular, the MMSE denoiser\cite{WeiYu, AMP_2} in the  $t$-th iteration can be expressed as:
\begin{gather}\label{Denoiser}
\eta(r_n^t,\sigma_h)=\frac{\alpha r_n^t}{1+\frac{1-\rho}{\rho}\beta\text{exp}(-\gamma|r_n^t|^2)},
\vspace{-5mm}
\end{gather}
where $\alpha=\frac{\sigma^2_h}{\sigma^2_h+\tau_t^2}$, $\beta=\frac{\sigma^2_h+\tau_t^2}{\tau_t^2}$ and $\gamma=\tau_t^{-2}-(\sigma^2_h+\tau_t^2)^{-1}$.
In addition, the first input of the MMSE denoiser $\mathbf{r}^t = [r_1^t, r_2^t,...r_N^t]^T$ can be interpreted as the matched filtered output $\mathbf{r}^t=\mathbf{P^*}\mathbf{\hat{x}}^{t} + \mathbf{x}^t$, which can be approximately modelled by the estimated signals $\mathbf{x}^t$, impaired by the AWGN.
Invoking the state evolution technique \cite{AMP_2}, it is derived by $\tau_{t+1}^2=\sigma_z^2 + \frac{N}{M}\mathbb{E}[|\eta(\mathbf{r}^t, \sigma_h)-\mathbf{x}|^2]$ with an initialization of $\tau_{0}^2=\sigma_z^2 + \frac{N}{M}\mathbb{E}[|x_n|^2]$.
In practice, an empirical estimation of $\tau_{t+1}^2$ can be computed by $\tau_{t+1}^2 = \frac{1}{M}||\mathbf{v}^t||^2_2$, where $||\cdot||_2^2$ denotes the square value of the $\ell_2$-norm.
In fact, the state evolution technique \cite{AMP_2} can characterize and predict the performance of the AMP algorithm during each iteration.
Interested readers are referred to \cite{AMP_2} for more details.

\setcounter{figure}{1}
\begin{figure*}[ptb!]
\centering
  \includegraphics[width=\textwidth-2cm,height=6cm]{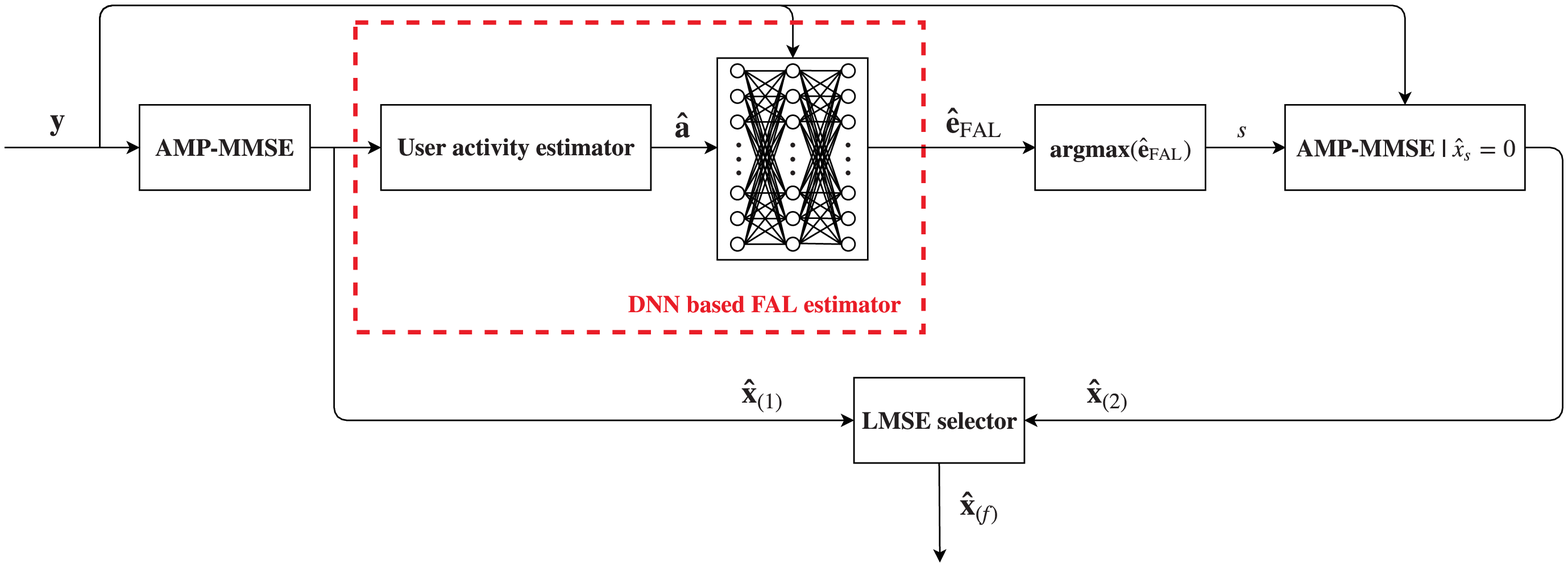}
  \caption{Flowchart for the proposed DL-LAMP algorithm.}
  \label{Flowchart}
\end{figure*}

\setcounter{figure}{0}
\begin{figure}[tb]
	\centering
	\includegraphics[width=80mm]{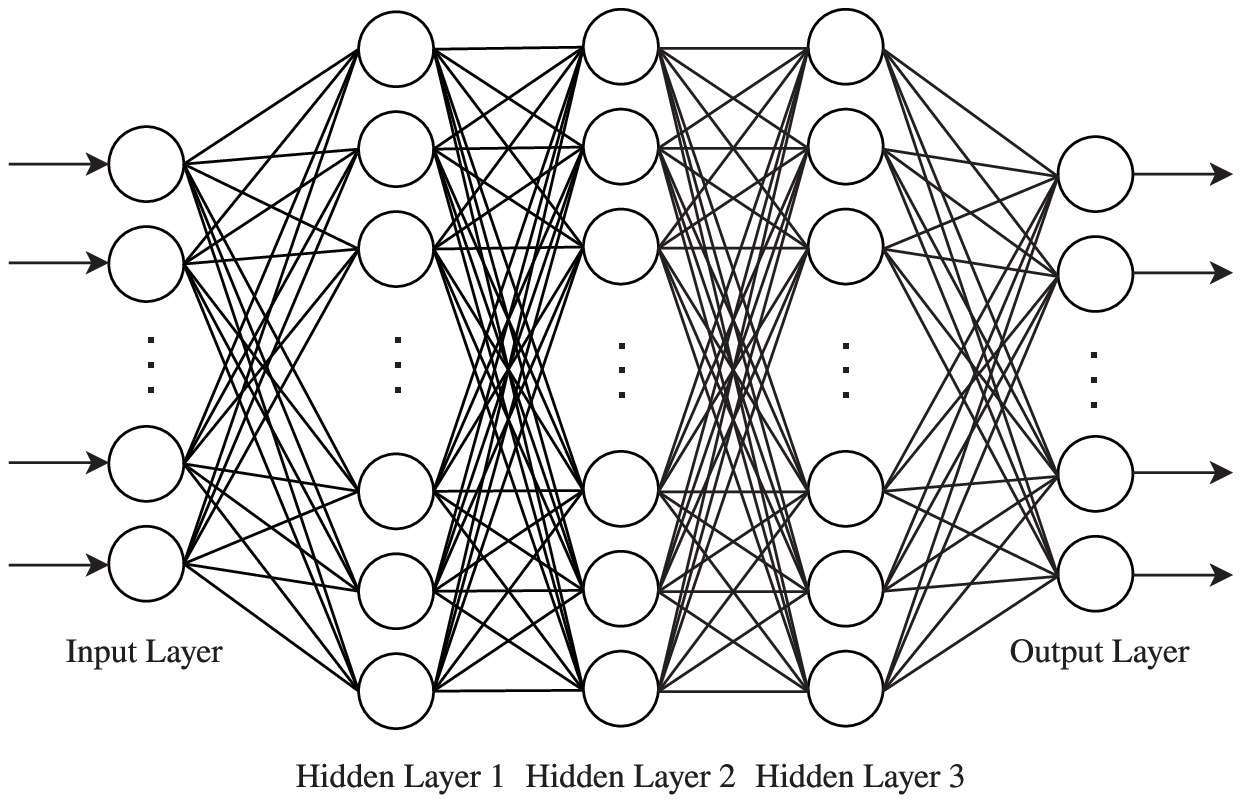}
    \vspace{-3mm}
	\caption{An example of DNN with 3 hidden layers.}
    \label{DNN}
\vspace{-2mm}
\end{figure}

\setcounter{figure}{2}

\subsection{Overview of Deep Neural Network}

In this section, we briefly describe the essentials of a deep neural network (DNN).
Fig. \ref{DNN} depicts an ordinary structure of a fully connected deep neural network, where the network mainly comprises of three types of neural layers.
The input layer takes the features to be processed and conveys the processed information to several hidden layers.
The outputs of each hidden layer are inputs to the following layer.
The last hidden layer is connected to the output layer which yields the network's output.
Each layer comprises of neurons such that each neuron computes an affine combination of its inputs, where the weights and bias of the affine combination are tunable parameters that are calibrated during the neural network ``learning'' process.
Activation functions such as hyperbolic tangent and sigmoid functions can be further applied in the neurons to improve the expressiveness beyond what is possible with only linear processing \cite{DL_Textbook}.

In the case of a supervised learning, the neural network usually contains three phases, including the training phase, validation phase and application phase.
During the training phase, a data set of input features and output labels (desired output corresponding to the neural network input) are fed into the neural network to adjust the values of the tunable parameters by backpropagating the losses between the labels and the neural network output.
The network's performance is evaluated using validation data set so as to select its parameters before over-fitting such as in early stopping, or to further refine the network's architecture.
Once the neural network is fully trained, the network is capable to realize the desired functionality to generate the corresponding output for any given input features in the application phase.

The capability of a neural network to approximate a fairly large family of input-output mappings stems from a flexible design of connections between neurons and selection of activation functions associated with them.
This motivates us to employ a neural network as a FAL estimator to learn the features of false alarm event during the AMP detection and the implicit correlation structure in the quantized pilot matrix.
%

\section{Proposed DL-LAMP algorithm}
Non-orthogonal pilot sequences cause inter-user interference which has a detrimental impact on user identification and channel estimation in mMTC. This is further exacerbated in the scenario of low-resolution quantization of pilot sequences.
To combat the interference, based on the observations and the estimated user activity, we propose to employ a neural network to find the suspicious device which is most likely to be falsely alarmed.
Then, the suspicious device is enforced to be inactive in each iteration of the second round of the AMP to avoid the interference for other devices.
In this section, we first describe the proposed DL-LAMP algorithm and introduce the training scheme of the neural network.

\subsection{Algorithm Structure}
According to the list decoding technique\cite{OSD}, generating a list of estimates $\mathbf{\hat{x}}$ by flipping some of the unknown variables is an effective method to further enhance the estimation performance.
In this paper, we generate two estimates $\mathbf{\hat{x}}_{(1)}$ and $\mathbf{\hat{x}}_{(2)}$ by executing the AMP for two rounds, as illustrated in Fig. \ref{Flowchart}.
In the first round, the observations $\mathbf{y}$ are fed to the AMP-MMSE processor to obtain the first estimate $\mathbf{\hat{x}}_{(1)}$.
Based on $\mathbf{\hat{x}}_{(1)}$ and $\mathbf{y}$, a DNN-based FAL estimator is proposed and designed for predicting the likelihood of each device to be falsely alarmed, i.e., wrongly detected as being active.
Then, the index of the suspicious device $s$ can be easily acquired via comparing the FALs of all the devices.
In the second round, the unknown variable associated with the suspicious device $s$ is set to be 0 in every iteration of the second round AMP-MMSE, i.e., $\hat{x}_s=0$ between Eqs. (\ref{x_hat}) and (\ref{v}).
Consequently, the suspicious device is forced to be inactive and the corresponding signal $\hat{x}_s$ is no longer interfering other users.
Once $\mathbf{\hat{x}}_{(1)}$ and $\mathbf{\hat{x}}_{(2)}$ are both estimated by the AMP-MMSE algorithm, the final estimate $\mathbf{\hat{x}}_{(f)}$ is chosen by a least mean squared error (LMSE) selector that
\begin{gather}\label{genie_aided}
\mathbf{\hat{x}}_{(f)} = \text{argmin}_{\hat{x}_i} |\mathbf{y}-\mathbf{P}\mathbf{\hat{x}}_{(i)}|^2, \text{where} \ i \in {1,2}.
\end{gather}

\subsection{Genie-aided List AMP}
To validate the effectiveness the idea of list decoding in improving the estimation performance of the AMP-MMSE algorithm, we employ a genie-aided selector to identify the suspicious device.
Note that the user activity $\mathbf{a} = \left[a_1,a_2,\ldots,a_N \right]^{{T}}$ and the signals $\mathbf{x}$ are known by the genie-aided selector.
Additionally, we note that this is identical to the training phase of a neural network where $\mathbf{x}$ and $\mathbf{a}$ are given.

For a genie-aided selector, a vector of element-wise Euclidean distance between $\mathbf{\hat{x}}_{(1)} = [\hat{x}_{(1),1},\hat{x}_{(1),2},\ldots,\hat{x}_{(1),N}]^{{T}}$ and $\mathbf{x}$ is $E(\mathbf{\hat{x}}_{(1)},\mathbf{x})=[|\hat{x}_{(1),1}-x_1|^2, |\hat{x}_{(1),2}-x_2|^2,...,|\hat{x}_{(1),N}-x_N|^2]^T$.
Define $(\sim\mathbf{a})$ as the vector which flips the value of 0 to 1 and 1 to 0 in the activity vector $\mathbf{a}$.
A vector $\mathbf{B} = (\sim\mathbf{a}) \bigodot E(\mathbf{\hat{x}},\mathbf{x}) = [B_1,B_2,\ldots,B_N]$ characterizes the element-wise errors of the inactive devices, where $\bigodot$ denotes the point-wise multiplication.
The larger the $B_n$, the more likely that inactive device $n$ will be falsely alarmed.
And $s=\text{argmax}(\mathbf{B})$ identifies the suspicious device who suffers the highest estimation error.
Then, in the second round of the AMP-MMSE estimation, the signal $\hat{x}_s$ is set to be 0 in each iteration update of $\mathbf{\hat{x}}$.

\begin{table}[t]
\centering
\caption{NMSE performance of AMP-MMSE and GA-LAMP.}
\begin{center}
 \begin{tabular}{c|c| c}
\hline
 Number of iterations & AMP-MMSE (dB) & GA-LAMP (dB) \\
 \hline
 3 &  -3.61 & -3.60 \\
  \hline
 5 & -4.55 & -4.80 \\
 \hline
 10 & -5.34 & -6.22  \\
 \hline
 20 & -5.67 & -6.99  \\
 \hline
\end{tabular}
\end{center}
\label{GA_LAMP}
\end{table}

Table \ref{GA_LAMP} shows the NMSE of AMP-MMSE and LAMP algorithm with genie-aided selector (GA-LAMP), respectively. In our simulations, the number of devices is $N = 150$ and the pilot length is $M = 30$.
The elements in the pilot matrix $\mathbf{P}$ is first drawn from i.i.d $\mathcal{CN}(0,\frac{1}{M})$, then quantized into 3-bit resolution.
The transmission probability is set to be $\rho = 0.1$ and the SNR is 40 dB.
It can be seen that for 20 iterations, the GA-LAMP provides a NMSE performance gain up to 1.32 dB compared to the conventional AMP-MMSE.
This confirms the effectiveness of our proposed list AMP algorithm structure and motivates us to construct a DNN to achieve the same functionality as the genie-aided selector.
\begin{algorithm}[t!]
  \caption{Deep Learning Assisted List AMP Algorithm}
        \begin{enumerate}[Step 1: ]
            \item Perform the first round AMP-MMSE to acquire the estimated sequence $\mathbf{\hat{x}}_{(1)}$.
            \item Prepare the inputs of the neural network by normalizing $\mathbf{y}$ and estimating $\mathbf{\hat{a}}$ as \eqref{hat_a}.
            \item Feed the preprocessed inputs to the neural network then find the index of suspicious device by $s=\text{argmax}({\mathbf{\hat{e}}}_{\rm{FAL}})$ based on the neural network's output.
            \item Perform the second round AMP-MMSE estimation by setting $\mathbf{v}^0=\mathbf{y}$ and $\mathbf{\hat{x}}^0=\mathbf{0}$, with the signal $\hat{x}_s$ saturated to be 0 in every update of $\mathbf{\hat{x}}^t$.
            \item Perform the LMSE selector to determine the final output $\mathbf{\hat{x}}_{(f)}$ according to \eqref{genie_aided}.
        \end{enumerate}
\end{algorithm}

\subsection{DNN-based FAL Estimator}





%
The attractiveness of employing the neural network is to investigate the features of the likelihood of the false alarmed devices in the training phase.
Then based on the ``learned'' experience, the neural network is capable to predict the most likely false alarmed device in the application phase.

\textbf{\underline{DNN Input Preprocessing:}}

In our model, the neural network is expected to estimate the FAL of each device based on the AMP estimates $\mathbf{\hat{x}}_{(1)}$ and the observation $\mathbf{y}$.
However, since each entry of the first round AMP output includes the channel estimates $\hat{h}_{(1),n}$ and the detected user's activity $\hat{a}_{(1),n}$, i.e., $\hat{x}_{(1),n} = \hat{a}_{(1),n} \hat{h}_{(1),n}$, $\mathbf{\hat{x}}_{(1)}$ can be interpreted as soft information while $\hat{a}_{(1),n}$ is the hard decision for user identification.
To facilitate the neural network training, we prefer to utilize the hard decision $\hat{a}_{(1),n}$ as its input rather than the soft information $\mathbf{\hat{x}}_{(1)}$.
In fact, the values of $\mathbf{\hat{x}}_{(1)}$ vary significantly with different realizations of noise $\mathbf{z}$ and desired unknown signals $\mathbf{x}$, which disrupts the training phase.
Therefore, we propose to transform the first round AMP estimates $\mathbf{\hat{x}}_{(1)}$ into the user activity sequence $\mathbf{\hat{a}} = [\hat{a}_1, \hat{a}_2,...,\hat{a}_N]^T$ by a user activity estimator as follow:
\begin{gather}\label{hat_a}
\hat{a}_n = \Lambda\bigg{(}|\hat{x}_n|, \delta = \xi\left(\mathbf{\hat{x}},\rho\right)\bigg{)},
\end{gather}
which returns 1 if $|\hat{x}_n| > \delta$ and 0, otherwise.
Its second argument $\delta$ denotes the detection threshold which is a percentile function \cite{Numpy2015} of the AMP estimates $\mathbf{\hat{x}}$ and the transmission probability $\rho$.
In particular, the percentile function $\xi\left(|\mathbf{\hat{x}}|,\rho\right)$ returns a threshold $\delta$ which leads to that only $\mathrm{Round} \left(\rho N\right)$ entries of $|\mathbf{\hat{x}}|$ are above $\delta$.
Here, $\mathrm{Round} \left(\cdot\right)$ returns the nearest integer of its input.

The threshold returned from $\xi\left(\mathbf{\hat{x}},\rho\right)$ splits all the sensors to $\mathrm{Round} \left(\rho N\right)$ active users and $N - \mathrm{Round} \left(\rho N\right)$ inactive users based on the AMP estimates $\mathbf{\hat{x}}$.
In the other words, the resulting user activity sequence $\mathbf{\hat{a}}$ has $\mathrm{Round} \left(\rho N\right)$ entries of 1 and $N - \mathrm{Round} \left(\rho N\right)$ entries of 0.
As a result, the input preprocessing roughly generates user activity estimates whose distribution is Pr$(\hat{a}_n=1)=\rho$ and Pr$(\hat{a}_n=0)=1-\rho$, $\forall n$, which is consistent with the prior user activity distribution.
%
%
%
%
%

\textbf{\underline{DNN Labels Preprocessing:}}

In the previous section, it has been shown that the vector $\mathbf{B}$ can be interpreted as the FAL of the devices.
Moreover, the estimate $\mathbf{\hat{x}}$ varies significantly with different channel realizations, where the vector $\mathbf{B}$ changes correspondingly.
In practice, to facilitate the training process of the neural network, we normalize each entry in the vector $\mathbf{B}$ to be within a range between 0 and 1.
Towards that end, the min-max normalization is applied to the vector $\mathbf{B}$.
Define $\mathbf{e}_{\rm{FAL}} = [e_{\rm{FAL},1},e_{\rm{FAL},2},\ldots,e_{\rm{FAL},N}] ^ T$ as the labels for the neural network,
which is the normalized vector of $\mathbf{B}$, i.e.,
\begin{gather}\label{hat_e}
e_{\rm{FAL},n} = \frac{B_n-\text{min}(\mathbf{B})}{\text{max}(\mathbf{B})-\text{min}(\mathbf{B})},
\end{gather}
where min($\cdot$) and max($\cdot$) find the minimum and maximum value of a vector, respectively.
Note that, in the application phase, the trained neural network outputs a sequence of soft information $\mathbf{\hat{e}}_{\rm{FAL}}$ on FAL.
This motivates us to employ the mean squared error as the loss function for training, i.e.,
\begin{align}\label{LossFunction}
  L(\mathbf{\hat{e}}_{\rm{FAL}}, \mathbf{e}_{\rm{FAL}}) &= \frac{1}{N}\sum_{n=1}^{N}|e_{\rm{FAL},n}-\hat{e}_{\rm{FAL},n}|^2,
\end{align}
A backpropagation optimizer to minimize the loss function $L(\mathbf{\hat{e}}_{\rm{FAL}}, \mathbf{e}_{\rm{FAL}})$ is employed to adjust the values of the tunable parameters in the neural network.

The overall DL-LAMP algorithm is summarized in Algorithm 1.
In Step 2, the observations are normalized to have zero mean and unit variance before feeding to the neural network to accelerate the training process.

\section{Simulation Results}
In this section, we evaluate the performance of the proposed DL-LAMP algorithm, and compare it with the AMP-MMSE.
In our simulations, we construct a simple DNN to demonstrate the concept of the proposed algorithm.
We consider a system of 150 devices that transmit pilot sequences of length 30.
The pilot matrix is chosen randomly by i.i.d $\mathcal{CN}(0,\frac{1}{M})$, then quantized to 3-bit resolution.
The fading coefficient for each device is randomly generated from i.i.d $\mathcal{CN}(0,1)$.
We employ the AMP-MMSE algorithm as proposed in \cite{WeiYu,AMP_2}.
The neural network is constructed and trained using Tensorflow.
The structure and the hyper-parameters of the DNN are listed in Table. \ref{table:2}.
Since $\mathbf{y}$ is sampled by complex values, the real and imaginary parts are formatted into corresponding two-column matrices and then simultaneously fed into the DNN.

\begin{table}[t]
\centering
\caption{The structure of DNN and the hyper-parameters for training.}
\begin{center}
 \begin{tabular}{c|c}
\hline
 Number of hidden layer & 2 \\
 \hline
 Hidden layer size &   $2 \times (M + N)$ \\
  \hline
 Hidden layer activation function &   Hyperbolic tangent \\
 \hline
 Output layer activation function &   Sigmoid \\
 \hline
 Optimizer & Root Mean Square Propagation \\
 \hline
 Learning rate & 0.001  \\
 \hline
 Batch size & 600  \\
 \hline
 AMP-MMSE iteration & 20 \\
 \hline

\end{tabular}
\end{center}
\vspace{-2mm}
\label{table:2}
\vspace{-4mm}
\end{table}

To save the training duration, the neural network is trained with the outputs of the AMP-MMSE after the 20-th iteration, while it is tested for different numbers of iterations.
For different SNR system setup, the neural network is trained individually.

The neural network is trained off-line and in the application phase, all the tunable parameters, the weights and biases in the fully connected neural layers, remained constant.
In our simulation, 2 hidden layers are employed in the fully connected neural network. In total, regarding the linear computations, there are $8M^2+8N^2+16MN$ multiplications and $4M+5N$ additions in the application phase.
Moreover, there are $4M+5N$ non-linear computations, where the non-linear function for each layer is stated in Table. \ref{table:2}. Note that the computations for the neurons at each layer are in parallel, so that the processing delay in the application stage is negligible.

\begin{figure}[t]
	\centering
	\includegraphics[width=85mm]{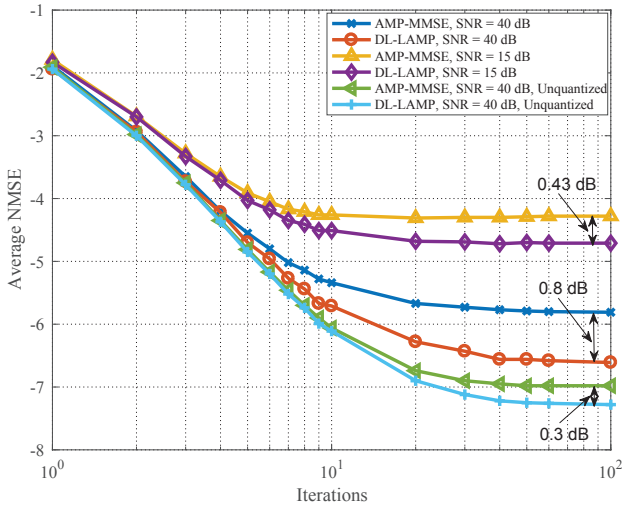}
	\caption{Average NMSE versus the number of iterations for DL-LAMP and AMP-MMSE.}
    \label{150_30_01}
\end{figure}

The simulated performance results are shown in Fig. \ref{150_30_01}, where AMP-MMSE indicates the AMP algorithm with MMSE denoiser, DL-LAMP refers to the proposed algorithm, the label ``Unquantized'' indicates that a pilot matrix is employed without quantization.
We can observe that, at the $100$-th iteration, the proposed DL-LAMP algorithm provides $0.8$ dB and $0.43$ dB NMSE gain compared to that at the AMP-MMSE at the SNR of $40$ dB and $15$ dB, respectively.
Moreover, at SNR = $40$ dB, compared to the unquantized case, it can be seen that the proposed algorithm provides more performance gain for the quantized pilot matrix, which is more practical for IoT sensors.
This relates to the fact that for a quantized pilot matrix, the pilot sequences of users are more correlated.
As a result, enforcing the suspicious device as inactive provides better performance.

\begin{figure}[t]
	\centering
	\includegraphics[width=85mm]{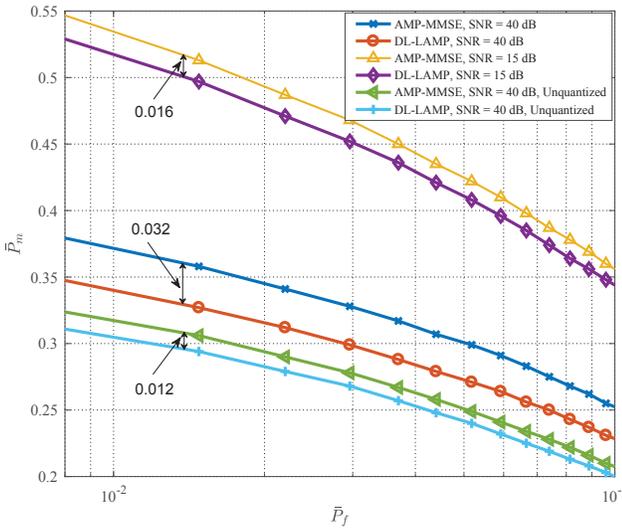}
	\caption{Missed detection probability versus false alarm probability for DL-LAMP and AMP-MMSE.}
    \label{P_m_P_f}
\end{figure}

We further investigate the corresponding performance of the average false alarm probability $\bar{P}_f$ versus the average missed detection probability $\bar{P}_m$, as shown in Fig. \ref{P_m_P_f}.
In particular, the false alarm and missed detection probability are defined as:
\begin{gather}\label{P_f_P_m}
P_f=\frac{\sum_{n=1}^{N}\mathbbm{1}\{\hat{a}_n=1, a_n=0\}}{\sum_{n=1}^{N}\mathbbm{1}\{a_n=0\}} \;\text{and}\\ P_m=\frac{\sum_{n=1}^{N}\mathbbm{1}\{\hat{a}_n=0,a_n=1\}}{\sum_{n=1}^{N}\mathbbm{1}\{a_n=1\}},
\end{gather}
respectively, where $\mathbbm{1}\{\cdot\}$ is the indicator function.
%
%
The average probabilities $\bar{P}_f$ and $\bar{P}_m$ are calculated by the mean value of $P_f$ and $P_m$ among all the realizations, respectively.
The AMP iteration is preset to be 100.
The false alarm probability is set between 0.1 and 0.008.
It is shown in Fig. (\ref{P_m_P_f}) that the proposed algorithm decreases the missed detection probability compared to the conventional AMP-MMSE algorithm, especially for the case with a quantized pilot matrix.

\section{Conclusion}
This paper proposed a DL-LAMP algorithm framework to improve the user identification performance in mMTC.
In particular, we construct a deep learning neural network, which serves as a FAL estimator.
Owing to benefit from the flexibility of the neural network design, the proposed DNN-based FAL estimator is able to learn the features of false alarm event during the AMP detection and the implicit correlation structure in the quantized pilot matrix.
Based on the FAL obtained from DNN, we identified a suspicious device which is most likely to be falsely alarmed.
Enjoying the benefit of list decoding, the proposed algorithm performs two rounds of AMP estimation, where in the second round, the suspicious device is enforced to be inactive.
Simulation results have shown that the proposed algorithm provides a performance gain and the corresponding missed detection probability is decreased.


\begin{thebibliography}{10}
\providecommand{\url}[1]{#1}
\csname url@samestyle\endcsname
\providecommand{\newblock}{\relax}
\providecommand{\bibinfo}[2]{#2}
\providecommand{\BIBentrySTDinterwordspacing}{\spaceskip=0pt\relax}
\providecommand{\BIBentryALTinterwordstretchfactor}{4}
\providecommand{\BIBentryALTinterwordspacing}{\spaceskip=\fontdimen2\font plus
\BIBentryALTinterwordstretchfactor\fontdimen3\font minus
  \fontdimen4\font\relax}
\providecommand{\BIBforeignlanguage}[2]{{%
\expandafter\ifx\csname l@#1\endcsname\relax
\typeout{** WARNING: IEEEtran.bst: No hyphenation pattern has been}%
\typeout{** loaded for the language `#1'. Using the pattern for}%
\typeout{** the default language instead.}%
\else
\language=\csname l@#1\endcsname
\fi
#2}}
\providecommand{\BIBdecl}{\relax}
\BIBdecl

\bibitem{3gpp}
``Study on new radio {(NR)} access technology physical layer aspects,'' 3GPP TR
  38.802, Tech. Rep., 2017.

\bibitem{Zhiqiang}
Z.~Wei, L.~Yang, D.~W.~K. Ng, J.~Yuan, and L.~Hanzo, ``On the performance gain
  of {NOMA} over {OMA} in uplink communication systems,'' \emph{arXiv preprint
  arXiv:1903.01683}, 2019.

\bibitem{WeiJSTSP2019}
Z.~{Wei}, D.~W.~K. {Ng}, and J.~{Yuan}, ``{NOMA} for hybrid mmwave
  communication systems with beamwidth control,'' \emph{IEEE J. Select. Topics
  Signal Process.}, vol.~13, no.~3, pp. 567--583, Jun. 2019.

\bibitem{SunPNC}
Z.~Sun, L.~Yang, J.~Yuan, and D.~W.~K. Ng, ``Physical-layer network coding
  based decoding scheme for random access,'' \emph{IEEE Trans. Veh. Technol.},
  vol.~68, no.~4, pp. 3550--3564, Apr. 2019.

\bibitem{Liu18ma}
L.~Liu, E.~G. Larsson, W.~Yu, P.~Popovski, C.~Stefanovic, and E.~de~Carvalho,
  ``Sparse signal processing for grant-free massive connectivity: A future
  paradigm for random access protocols in the {I}nternet of {T}hings,''
  \emph{IEEE Signal Process. Mag.}, vol.~35, no.~5, pp. 88--99, Sep. 2018.

\bibitem{RAbbas18}
R.~{Abbas}, M.~{Shirvanimoghaddam}, Y.~{Li}, and B.~{Vucetic}, ``A novel
  analytical framework for massive grant-free {NOMA},'' \emph{IEEE Trans.
  Commun.}, vol.~67, no.~3, pp. 2436--2449, Mar. 2019.

\bibitem{SunJUICE}
Z.~Sun, Z.~Wei, L.~Yang, J.~Yuan, X.~Cheng, and L.~Wan, ``Joint user
  identification and channel estimation in massive connectivity with
  transmission control,'' in \emph{Proc. IEEE Intern. Sympos. on Turbo Codes
  Iterative Information Process.}, 2018, pp. 1--5.

\bibitem{SunJUICE2019}
Z.~{Sun}, Z.~{Wei}, L.~{Yang}, J.~{Yuan}, X.~{Cheng}, and L.~{Wan},
  ``Exploiting transmission control for joint user identification and channel
  estimation in massive connectivity,'' \emph{IEEE Trans. Commun.}, early
  access, 2019.

\bibitem{WeiYu}
Z.~Chen, F.~Sohrabi, and W.~Yu, ``Sparse activity detection for massive
  connectivity,'' \emph{IEEE Trans. Signal Process.}, vol.~66, no.~7, pp.
  1890--1904, Apr. 2018.

\bibitem{Liu18}
L.~Liu and W.~Yu, ``Massive connectivity with massive {MIMO}-part {I}: Device
  activity detection and channel estimation,'' \emph{IEEE Trans. Signal
  Process.}, vol.~66, no.~11, pp. 2933--2946, Jun. 2018.

\bibitem{lowADC}
J.~{Zhang}, L.~{Dai}, S.~{Sun}, and Z.~{Wang}, ``On the spectral efficiency of
  massive {MIMO} systems with low-resolution {ADCs},'' \emph{IEEE Commun.
  Lett.}, vol.~20, no.~5, pp. 842--845, May. 2016.

\bibitem{Milutin2019}
M.~Pajovic, T.~Koike-Akino, and P.~V. Orlik, ``Model-driven deep learning
  method for jammer suppression in massive connectivity systems,'' \emph{arXiv
  preprint arXiv:1903.06266}, 2019.

\bibitem{DLAMP_Philip2016}
M.~{Borgerding} and P.~{Schniter}, ``Onsager-corrected deep learning for sparse
  linear inverse problems,'' in \emph{Proc.IEEE Global Conf. on Signal and Inf.
  Process.}, Dec. 2016, pp. 227--231.

\bibitem{DLISTA}
U.~S. {Kamilov} and H.~{Mansour}, ``Learning optimal nonlinearities for
  iterative thresholding algorithms,'' \emph{IEEE Signal Process. Lett.},
  vol.~23, no.~5, pp. 747--751, May 2016.

\bibitem{DLFISTA}
K.~Gregor and Y.~LeCun, ``Learning fast approximations of sparse coding,'' in
  \emph{Proc. Int. Conf. on Machine Learning}, 2010, pp. 399--406.

\bibitem{DLAMP}
H.~{He}, C.~{Wen}, S.~{Jin}, and G.~Y. {Li}, ``Deep learning-based channel
  estimation for beamspace mmwave massive \text{MIMO} systems,'' \emph{IEEE
  Wireless Commun. Lett.}, vol.~7, no.~5, pp. 852--855, Oct. 2018.

\bibitem{DBAMP}
C.~A. {Metzler}, A.~{Maleki}, and R.~G. {Baraniuk}, ``From denoising to
  compressed sensing,'' \emph{IEEE Trans. Inf. Theory}, vol.~62, no.~9, pp.
  5117--5144, Sep. 2016.

\bibitem{OSD}
M.~Fossorier and S.~Lin, ``Soft-decision decoding of linear block codes based
  on ordered statistics,'' \emph{IEEE Trans. Inf. Theory}, vol.~41, pp. 1379 --
  1396, 10 1995.

\bibitem{threesigma}
E.~Grafarend, \emph{Linear and Nonlinear Models: Fixed Effects, Random Effects,
  and Mixed Models}.\hskip 1em plus 0.5em minus 0.4em\relax Walter de Gruyter,
  2006.

\bibitem{Quantization}
R.~M. Gray and D.~L. Neuhoff, ``Quantization,'' \emph{IEEE Trans. Inf. Theory},
  vol.~44, no.~6, pp. 2325--2383, Sep. 2006.

\bibitem{AMP}
D.~L. Donoho, A.~Maleki, and A.~Montanari, ``Message-passing algorithms for
  compressed sensing,'' \emph{Proc. of the National Academy of Sciences}, vol.
  106, no.~45, pp. 18\,914--18\,919, 2009.

\bibitem{STLASSO}
A.~{Maleki}, L.~{Anitori}, Z.~{Yang}, and R.~G. {Baraniuk}, ``Asymptotic
  analysis of complex {LASSO} via complex approximate message passing
  {(CAMP)},'' \emph{IEEE Trans. Inf. Theory}, vol.~59, no.~7, pp. 4290--4308,
  Jul. 2013.

\bibitem{AMP_2}
D.~L. {Donoho}, A.~{Maleki}, and A.~{Montanari}, ``Message passing algorithms
  for compressed sensing: {I.} motivation and construction,'' in \emph{IEEE
  Inf. Theory Workshop on Inf. Theory}, Jan. 2010, pp. 1--5.

\bibitem{DL_Textbook}
I.~Goodfellow, Y.~Bengio, and A.~Courville, \emph{Deep Learning}.\hskip 1em
  plus 0.5em minus 0.4em\relax The MIT Press, 2016.

\bibitem{Numpy2015}
T.~E. Oliphant, \emph{Guide to NumPy}, 2nd~ed.\hskip 1em plus 0.5em minus
  0.4em\relax USA: CreateSpace Independent Publishing Platform, 2015.

\end{thebibliography}


\end{document}